Another Dirac
Jeremy Bernstein
Stevens Institute of Technology

"A description of the proposed formulation of quantum mechanics might best begin by recalling some remarks made by Dirac concerning the analogue of the Lagrangian and the action in quantum mechanics. These remarks bear so directly on what is to follow and are so necessary for an understanding of it, that it is thought best to quote them in full even though it results in a rather long quotation." Richard Feynman[1]

"This note is concerned with methods of separating isotopes which depend on subjecting the mixture of isotopes, in liquid or gaseous form, to physical conditions which tend to cause a gradient in the concentration of an isotope. The most useful examples of such physical conditions are the presence of a field of centrifugal force or of a temperature gradient. There is a general theory governing the performance of a separator which employs such a method. It puts an upper limit to the output of the apparatus and shows what running conditions one should strive to attain in order to approach the theoretical limit in practice." P.A.M. Dirac[2]

I: Paths

I would imagine that if the average physicist was asked to list Dirac's achievements in physics he or she would say "the Dirac equation." A few might say "the Dirac delta function" but this is a mathematical convenience and not exactly a discovery in physics.[3] Some might also say

---

[1] Feynman's Thesis; A New Approach to Quantum Theory, edited by Laurie M. Brown, World Scientific, New Jersey, 2005, p.26
[2] This paper, "The Theory of the Separation of Isotopes by Statistical Methods," was never published by Dirac but is held in the Public Records Office in Kew, London. I am grateful to Helmut Rechenberg for supplying a copy.
[3] Dirac introduced it in a paper entitled The Physical Interpretation of the Quantum Dynamics, Proc. R.Soc,Lond.A 1927,**113**,621-641.

Fermi-Dirac statistics but then wonder why the alphabetical order of the names had been reversed. Dirac himself made this clear when he wrote in his quantum mechanics text, " This [ the Pauli principle] is an important characteristics of particles for which  only antisymmetrical states occur in nature. It leads to a special statistics which was first studied by Fermi, so we shall call particles for which only antisymmetrical states occur in nature *fermions*."[4] Of course Dirac made important contributions to the formulation of quantum theory, some of which I will shortly discuss, but in awarding him the Nobel Prize in 1933-which he shared with Schrödinger-these were not discussed least of all by Dirac. His Nobel lecture was entirely concerned with the Dirac equation and anti-matter. I believe that if Carl  Anderson had not discovered the positron in 1932 Dirac would not have gotten the prize at that time. In this note I am going to discuss two of Dirac's contributions that for one reason or another are not discussed as frequently. The first will be of an obscure paper in which Dirac presented the ideas that led to the path integral formulation of quantum mechanics which was first exploited in Richard Feynman's PhD thesis. The second  will be Dirac's theory of the separation of isotopes using the gas centrifuge. This work was done at the beginning of the Second World War and introduced ideas that have been the basis of this subject ever since.

I had one opportunity to ask Dirac about the origins of the quantum theory. This was many years ago but I have a vivid memory of his emphasis on Heisenberg. As far as Dirac was concerned it was Heisenberg who made the decisive step that liberated physicists from the old quantum theory which was an uneasy mixture of classical and quantum physics. In 1925, when Dirac was a graduate student at Cambridge, Heisenberg spoke at the university. The venue was the so-called "Kapitza Club". Pytor Kapitza was a Russian physicist who spent a decade in Cambridge, even directing his own laboratory. In 1934 he visited the Soviet Union. His passport was removed and he was not allowed to leave. He spent the rest of his career there. In Cambridge he founded this club for the purpose of sharing the latest work in physics. Membership was by invitation. Dirac was not sure if at the time of Heisenberg's lecture he had been invited to join, but he was sure that he had not heard the lecture. His tutor Ralph Fowler had a proof of Heisenberg's paper which he sent to Dirac who was then in Bristol visiting his parents. On the paper Fowler scrawled a note asking for Dirac's comments. The paper was in German but Dirac knew enough German to be able to read it. When he read it, it reminded Dirac of something he remembered from classical mechanics-the Poisson brackets. But he was not quite sure what the definition of these brackets was so he had to wait until his return to Cambridge where he could consult a book.

---

[4] Quantum Mechanics, by  P.A.M. Dirac, Oxford University Press, Oxford,  1947, p.210.

Heisenberg states his program at the beginning of his paper," It is well-known that the formal rules which are used in quantum theory for calculating observable quantities such as the energy of the hydrogen atom may be seriously criticized on the grounds that they contain as basic element, the relationships between quantities that are apparently unobservable in principle, e.g. position and period of the revolution of the electron. Thus these rules lack an evident physical foundation, unless one still wants to retain the hope that the hitherto unobservable quantities may later come within the realm of experience…"[5] Heisenberg expands two position functions $x(t)$ and $y(t)$ in Fourier series and notes that if the terms in the sum obey the quantum mechanical combination rules the $x$ and $y$ do not commute. We realize looking at his paper that Heisenberg is doing matrix multiplication something that he did not realize when he wrote it. He seems almost embarrassed by the failure of commutivity and refers to it as a "difficulty." What you will not find in the paper is the canonical commutation rule $qp-pq=i\hbar I$ where $I$ is the identity. Still less will you find the "Heisenberg equation" $A(t)H-HA(t)=i\hbar d/dt A(t)$ where $A$ is some operator and $H$ is the Hamiltonian. These equations were arrived at later independently by Dirac and Max Born and his student Pascual Jordan.

Dirac's paper, "The Fundamental Equations of Quantum Mechanics which was rushed through the process by Fowler was published in 1926.[6] has a fundamental assumption. If x and y are functions of q and p, coordinates and momenta which are represented by operators, then

$$xy-yx=i\hbar\{\partial x/\partial q\, \partial y/\partial p - \partial x/\partial p\, \partial y/\partial q\}\ .$$

(1)

---

[5] Quantum Theoretical Re-interpretation of Kinematic and Mechanical Relations, W.Heisenberg in Sources of Quantum Mechanics, edited by B.L. Van der Waerden, Dover Books, New York,1968.,261.

The quantity in the curly bracket is the Poisson bracket. From this the canonical commutation relation for p and q follows at once. By analogy with the Poisson bracket equation of motion Dirac also writes down the "Heisenberg equation." Dirac also discusses replacing p and q by what would later be called creation and annihilation operators but he does not take this discussion very far. The Born-Jordan paper, "On Quantum Mechanics"[7] is a much deeper paper than Dirac's. This is perhaps not too surprising. While Dirac was essentially a student Born was one of the most accomplished physicists of his time. Jordan was an extremely gifted physicist who has probably received less than his due because of his later Nazi associations. One of the things that Jordan did was to arrive at Fermi-Dirac statistics at about the same time that Fermi did. Born took Jordan's paper to the United States and then forgot to look at it so it was not published. Dirac's paper is certainly of historical interest but you could teach a significant part of a quantum mechanics course from that of Born and Jordan and that of the follow up paper-the "three man paper"-of which Heisenberg was a co-author. Born knew about matrix algebra and the first part his paper with Jordan is a tutorial in the subject. This is followed by a section in which the Heisenberg equation is derived.

The Lagrangian enters the discussion through the action

$$S - \int_{t_o}^{t_1} L\,dt = \int_{t_o}^{t_1} (p\dot{q} - H(p,q))\,dt\ .$$

They derive the Heisenberg equation from the assumption of what was later known as the "expectation value" be an extremal. As an application of the formalism they compute the energies of the anharmonic oscillator, The "three man paper" takes the formalism much further and includes a discussion of perturbation theory. It also has a brief discussion of transformation theory to which I want to turn next.

---

[6] Proc.Roy. Soc. A,**109** (1926) 642-653/
[7] See Van der Wareden op cit for a translation.

It appears that of the quantum mechanics papers, the 1927 paper to which I referred earlier was Dirac's favorite. Its elegant mathematics appealed to him greatly. Not only did he introduce the delta function but also the terminology "c-number" and "q-number" to distinguish between classical ordinary numbers and quantum mechanical operators. He introduced the notation ($\xi'/\eta'$) to represent matrix elements. The purpose of this paper was to show how the notion of the canonical transformations of classical Hamiltonian dynamics is realized in the quantum theory. It will be recalled that in classical mechanics it is sometimes useful to replace the coordinates q(t) and the momenta p(t) by new coordinates Q(q,p,t) and P(q.p.t) such that Hamilton's equations, where H is the Hamiltonian,

$$dp/dt = -\partial H/\partial q \qquad (2)$$

and

$$dq/dt = \partial H/\partial p \qquad (3)$$

are preserved. This imposes special conditions on the transformation which need not concern us here. The quantum mechanical version of these canonical transformations is a transformation of the operators. Using Dirac's notation. if g is the operator being transformed and G the result and b the operator generating the transformation then

$$G = bgb^{-1}. \qquad (4)$$

This transformation preserves the canonical commutation relations. If g is hermitian and we want to preserve this feature then we require that $b^{-1}=b^\dagger$ where $b^\dagger$ is the hermitian conjugate to b. Dirac does not in this paper use the term "unitary" for this type of transformation. Dirac wants to explore the unitary transformations that diagonalize g. If g is the Hamiltonian then these diagonal elements are the allowed energies of the system.

He considers two operators with matrix elements

$$\xi(\xi'\xi'') = \xi'\delta(\xi'-\xi'')$$

(5)

and

$$\eta(\xi'\varsigma") = -i\hbar \partial/\partial\varsigma' \delta(\xi'-\xi") .$$

(6)

One readily shows that these operators are canonically conjugate. Dirac then considers any function of these operstors, F(ξ,η). One wants to find a canonical transformation that diagonalizes this function. That is we want to reduce it to the form

$$F(α'α")=δ(α'-α")F(α').$$

(7)

In other words we need the matrix elements (ξ'/α') that accomplish the transformation. Dirac shows that these matrix elements obey the ordinary differential equation,

$$F(\xi',-i\hbar \partial/\partial \xi')(\xi'/\alpha') = F(\alpha')(\xi'/\alpha').$$

(8)

The F(α') are the diagonal matrix elements. He then notes that if the variables are identified with the coordinates and momenta q and p, and if F is the Hamiltonian, then the above equation is the time independent Schrödinger equation. Here the F(α') are the energies and the(ξ'/α') are the Schrödinger wave functions which in this view of things diagonalize the Hamiltonian. This rather straightforward argument demonstrates that the Schrödinger and Heisenberg pictures are simply two different representations of the same theory. This had already been claimed in a long and rather obscure paper by Schrödinger. [8] Curiously Dirac makes no reference to this paper although he does refer to a Schrödinger paper that had been published later. One wonders if he read the earlier paper or if he decided that it was irrelevant. His own argument is a masterpiece of economy. Jordan also discussed the transformation theory but his notations are pretty opaque compared to Dirac's.

I think that a fair summary of Dirac's work on the quantum theory to this point is that while it is very impressive,with the exception of the introduction of the Poisson brackets, it was work that in one form or another was also done by others. The "Dirac equation" which Dirac formulated in 1928 is something else. It is a work of inspired originality and it is for this that Dirac won the Nobel Prize. It is the same kind of originality that characterizes Dirac's work on the Lagrangian in quantum mechanics although it took some time for it to be appreciated. The reason for this was partly the odd way it was published which goes back to Dirac's nature. He was not a person who needed a great many human contacts. But like many solitary people those he had ran deep. One of them was with the aforementioned Russian physicist Kapitza. Prior to the publishing of his paper in 1933 Dirac had made some visits to the Soviet Union and had even done some mountain climbing there. The only sport that Dirac had any interest in was rock climbing. Thus Dirac chose to publish his paper, " The Lagrangian in Quantum Mechanics" in the now long defunct Russian journal ,*Physikalische Zeitschrift der Sowjetunion*[9] . This practically guaranteed that the paper would not be widely read. But Dirac published the basic ideas in his quantum mechanics text. However they are buried in the middle of the book and easily skipped over. In 1941 Feynman was looking for a way to quantize theories where there was no classical Hamiltonian. At this time Herbert Jehle was a visitor to Princeton and he called Feynman's attention to Dirac's work. Since Feynman never cited Dirac's paper but only the book my guess is that he never read the paper.

---

[8] Annalen der Physik, 4,**79**, 1926, 734-756

Dirac begins his paper by explaining why it was natural to quantize classical theories using Hamiltonian dynamics. Once you know how to represent position and momentum as operators it is elementary to represent any function of them such as the Hamiltonian as an operator. But this limits ones options. The Hamiltonian is not a relativistic invariant so this formulation is intrinsically non-relativistic. On the other hand the action S is a relativistic invariant so if it could be used one broadens ones possibilities. The key idea is the observation that going from a basis in which the coordinate q(t) is diagonal to one in which q at a different time, say T, is diagonal can be achieved by a canonical transformation in which the generating function is the action. Dirac makes the absolutely remarkable statement;

---

[9] Band3, Heft **1** (19330 PP.64-72.. It is reproduced in Brown op.cit.

" $\langle q_t | q_T \rangle$ corresponds to $\exp\left[i\int_T^t Ldt/\hbar\right]$ ".

What can "corresponds to" possibly mean? He does not explain either in his paper or his book. Feynman took this to mean that one must be proportional to the other and worked out the proportionality factors in some examples. He once asked Dirac if he had ever done the same and was told "no."

Dirac now considers the case where T and t differ from each other only infinitesimally and he makes a similar statement,

" $\langle q_{t+dt} | q_t \rangle$ corresponds to $\exp[[iLdt/\hbar]$ ".

Again there is no explanation of what "corresponds to" means. In ordinary quantum mechanics if one writes the solution to the time dependant Schrödinger equation as Aexp(i/ℏS) then under reasonable assumptions S obeys the Hamilton-Jacobi equation ∂S/∂t=-H(q,∂S/∂q) where H is the Hamiltonian. Thus S is the action. Dirac's relationship is a generalization of this result as he notes in his paper.

Dirac now imagines dividing the time interval between T and t into many short intervals. He then chooses "paths" between the q's at these times. He writes.

$$\langle q_t | q_T \rangle = \int \langle q_t | q_m \rangle dq_m \langle q_m | q_{m-1} \rangle dq_{m-1} ... \langle q_2 | q_1 \rangle . dq_1 \langle q_1 | q_T \rangle \qquad (9)$$

where the integral is a multiple integral over all the intermediate q's;ie a "path integral." Each of these scalar products will take the form that Dirac proposed in terms of an exponential of the action. Then Dirac asks what is the classical limit of this expression? Each of the scalar products involves the integral of the Lagrangian divided by ℏ. In the classical limit ℏ goes to zero. The integrands wildly oscillate with the exception of those paths in which the action is stationary.

But these are the classical paths had these are the ones that contribute to the expression in that limit. He closes his paper with some general remarks. One looks in vain for any application.

Feynman never published his thesis. Soon after he wrote it he and his advisor John Wheeler went off to war, After the war Feynman published some of the results in an article in *The Reviews of Modern Physics,*[10]"*Space-Time Approach to Non-Relativistic Quantum Mechanics*". In an appendix I will present a calculation of this probability amplitude in the simplest case where there is no interaction. Other cases like the harmonic oscillator can be done but more generally it is a difficult method to apply. Nonetheless the formalism and its descendants are at the heart of modern discussions of the quantum theory whether in the guise of "paths" or "worlds" or "histories." It is interesting to reflect that all of this can be traced back to an obscure paper by Dirac.

II: Separations

Dirac's interest in the separation of isotopes went back to the early 1930's. It was an active subject in Cambridge where Francis Aston, the inventor of the mass spectrograph, for which he won the 1922 Nobel Prize in chemistry, was a professor. Dirac proposed a method the generic name for which is a "stationary centrifuge." Here the gas to be separated into its isotopes moves while the object that does the separation remains stationary. Dirac's idea was to force the gas to move through a large angle in a bent tube. The heavy component would be bent less. He actually carried out an experiment much to the amusement of Rutherford. The results were hard to interpret. He was going to carry out more with Kapitza, but Kapitza was detained in Russia and Dirac dropped the matter. It was taken up again during the Second World War and the South Africans used a version of the stationary centrifuge to separate enough uranium isotopes to make several nuclear weapons which they destroyed without testing them in the 1990's.

---

[10] *Rev.Mod.Phys.* **20**,367-387 (1948) Both this paper and the thesis are included in Brown op.cit.

In March of 1940 Rudolf Peierls and Otto Frisch produced the memoranda that started the British nuclear weapons program and to a certain extent ours. It was immediately clear that the *sine qua non* was the separation of uranium isotopes. Dirac was contacted and he began his wartime activity devoted to isotope separation. It seems that it was in 1941 that Dirac wrote his seminal paper "The Theory of the Separation of Isotopes by Statistical Methods." This paper got to Peierls who was working at that time with Klaus Fuchs who even then was spying for the Russians. Peierls and Fuchs produced the standard paper on isotope separation [11] which found its way into the American work at the hands of people like Karl Cohen. These people all credit Dirac with the basic ideas.

Before I describe some of Dirac's contributions to the theory of isotope separations by centrifuges-especially gas centrifuges I need to describe briefly how such centrifuges work. They consist of cylinders some ten to twelve centimeters in diameter and a meter or two in length. Because the details of modern centrifuges are classified one cannot get precise specifications. The best modern centrifuges are made of carbon fiber and can rotate around their long axes at peripheral speeds of some 700 meters per second. Before the centrifuges begin to rotate that are put under vacuum to eliminate air resistance. Once they are rotating the gas is introduced. For the separation of the uranium isotopes U-235 and U-238 the gas that is used is uranium hexafluoride. The gas acquires the rotating motion of the cylinders. The heavier isotope is pushed more readily to the centrifuge wall by the centrifugal force. This is how the separation is produced. An obvious question to ask is in this case, won't the isotopes become completely separated if you leave the gas in long enough. To see why this is not the case an analogy is useful; the gravitational separation of isotopes in the stratosphere.

---

[11] "Separation of Isotopes" by K.Fuchs and R.Peierls, in Selected Papers of Sir Rudolf Peierls edited by R.H.Dalitz and Sir Rudolf Peiels, World Scientific, Singapore,1997,303-320

Consider a rectangular slab of atmosphere. If its total mass is m then a gravitational force of mg is pulling it down towards the earth's surface. But due to the difference in pressures at the top and bottom of the slab assuming that the density falls off as the distance above the earth's surface increases, there is a net upward pressure which balances the force of gravity. The difference in pressure is at equilibrium equal to the downward gravitational force;ie,

$dp = -g\rho dz$ where $\rho$ is the density of the gas and z is the height above the earth. If we assume that the gas is ideal and that the temperature remains the same throughout the slab-something that is not actually true for the atmosphere- we have the equation, with $\mu$ the molar mass and R the gas constant.

$$dp = -g\mu p/RT dz. \qquad (10)$$

or

$$dp/p = -g\mu/RT dz \qquad (11)$$

which integrates to

$$p/p_o = \exp{-((g\mu/RT)z)} \qquad (12)$$

which is the 'barometric formula".

In 1919 Frederick Lindemann and Francis Ashton published their seminal paper entitled " The Possibility of Separating Isotopes."[12] Ashton, as I have mentioned, won the 1922 Nobel Prize in Chemistry for his use of the mass spectrograph to separate isotopes and Lindemann became Churchill's war-time science advisor and later Lord Cherwell. In this paper they explore various separation methods one of which is the use of gravitation. They say that starting at a certain height above the earth, $h_o$, the isotopes of say neon, which is the case they study, will no longer mix by convection so that can be separated gravitationally. If you call the density of the heavy isotope $\rho_1$ and the density of the light one $\rho_2$, then with the assumptions that led to the barometric formula

$$\rho_1/\rho_2 = (\rho_1/\rho_2)_0 \exp(-(g\ \Delta h/RT(\mu_1 - \mu_2)).$$

(13)

---

[12] *Phil.Mag.*, Vol.xxxvii, p.523, 1919

Here Δh is the height above where the convection mixing stops. Aston and Lindemann suggest designing a balloon that would rise to 100,000 feet where it could sample the ambient atmosphere and look for second isotopes of neon. But they conclude, "Although the quantities are measurable they do not appear sufficiently striking to warrant the outlay and labour such experiments would entail."[13] There do seem to be some experiments on South Polar ice that show evidence for this kind of gravitational separation. More relevant to us is what Aston and Lindemann have to say about centrifuges.

They argue that the same equation holds if you substitute for the gravitational acceleration g the centrifugal acceleration $v^2/r=\omega^2 r$; At the edge of the centrifuge the ratio of densities would be

$$K/K_o = \exp(-v^2/RT(\mu_1 - \mu_2)). \qquad (14)$$

Here v is the peripheral velocity and K is the density ratio at the edge while $K_0$ is the density ratio at the center. They put in some numbers and concluded that a peripheral velocity of at least a thousand meters a second would be required to make useful separations. In 1934 J.W.Beams and F.R.Haynes used a centrifuge to separate the isotopes of gaseous chlorine. It had a maximum peripheral velocity of some 800 meters a second before it burst, Commercial gas centrifuges with modern materials can run at these speeds.

Given the assumptions, this formula is a useful way of estimating the percentage separation that a given centrifuge can perform. What it does not tell us is the rate at which this can be done. A theoretical maximum was supplied by Dirac in his paper. One should think of this the way one thinks of the Carnot cycle. The Carnot cycle provides the optimal performance of an ideal heat engine. Real heat engines will perform less well as will real centrifuges. Elsewhere I have presented a derivation of this maximum.[14] The result is for a cylindrical centrifuge of length h is in kilograms per second

---

$$U^{max} = \pi/2 D\rho(\Delta\mu/2RT)^2 h v^4.$$

(15)

Here ρ is the density of say the light component of the gas and v is the peripheral velocity and D is the diffusion coefficient measured in meters squared per second. The dependence on the fourth power is very striking but for actual centrifuges it is more like the power. My concern here however is Dirac's introduction of the unit he called "sep-power." This is a measure of how much separation power is needed to perform some given task such as producing a kilogram of 90% enriched uranium starting with uranium hexafluoride which used natural uranium with a concentration of about 0.7% uranium-235. The Dirac "value function" is used to calibrate this effort.

I am not going to present Dirac's derivation-at least as it is presented in the available part of his paper-since it assumes that the isotopic concentrations are all small. The paper refers to an appendix which has never been made available where this restriction is dropped. Rather I will present the derivation given by Peierls and Fuchs in 1942. They refer to Dirac's paper and presumably they saw the appendix.

When there is isotope separation there is an entropy change ΔS. Peierls and Fuchs label the concentration of the light isotope c and therefore the concentration of the heavy isotope is 1-c. They define an quantity F as

$$F = \Delta S/c(1-c).$$

(15)

In a footnote the explain the denominator. "The reason for this is that, with all usual methods, the work done by the device on the molecules is approximately independent of their nature. Of all possible pairs of molecules only the fraction c(1-c) are unlike ones, and only on those cases can the work be done for the purpose of distinguishing them lead to any discernible result. In all other cases it is wasted, Hence the factor c(1-c) in the efficiency."[15] They define a quantity

$$\Delta Y/\Delta t = R\Delta F/\Delta t \qquad (16)$$

as the separating power, To find this we need an expression for the change in entropy $\Delta S$ produced by a separation of two constituents in a binary mixture. $\Delta S$ is given by

$$\Delta S = -R(c\ln(c)+(1-c)\ln(1-c)) \qquad (17)$$

If you introduced a semi-permeable membrane into the original volume and, if there was a fifty-fifty admixture of the two components -c=1/2- then, after complete separation, the total entropy of the separated components would be

$$S = k\ln(2^N) \qquad (18)$$

where N is the total number of molecules. If you suppose a small change in the concentration, 'd' and expand in a Taylor series, this would produce a small change in the entropy, $\delta S$, given by

$$\delta S \sim Rd^2/2c(1-c), \qquad (19)$$

or

$$\Delta Y = Rd^2/2(c(1-c))^2. \qquad (20)$$

They introduce a quantity y(c) which represents a measure of the total effort to produce one mole of concentration c from and ordinary mixture of isotopes. The dimensionless quantity they define as the "separation potential." $\Delta Y$ can be expanded in a Taylor series and because of conservation laws the first non-zero term is the coefficient of the second derivative of y with respect to c. Hence cancelling terms one is led to the differential equation

$$d^2/dc^2 y(c) = 1/(2c^2(1-c)^2)$$

(21)

which has the solution

$$y(c) = (2c-1)\log(c/(1-c)) + ac + b$$

(22)

---

[15] Fuchs " Separation of Isotopes" op cit p.303.

where a and b the integration constants. They, following Dirac, fix these constants by insisting that if $c_0$ is the concentration of one of the components of the natural mixture then both y and its derivative must vanish at $c=c_0$. This gives them a form of the function

$$V=(2c-1)\ln((c/1-c))(1-c_c)/c_0))+(c-c_0)(1-2c_0)/c_0(1-c_0)$$

(24)

However the common treatment sets a=b=o. This leads to what is called the "Dirac value function" V(c) where

$$V(c)=(2c-1)\log(c/(1-c)).$$

(25)

To see how this is used I am going to consider the case of the separation of uranium isotopes by centrifuge. A gas centrifuge has a portal for the feed and two portals for the output. Through one of these output portals passes the "product"-the enriched uranium. Through the other passes the low enriched uranium or the "tails." In 1939 Harold Urey invented the idea of "counter currents." The heavier gas is made to move downward at the periphery while the light gas moves upward at the center. As one of his contributions Dirac worked out the basic theory of this which was the foundation of the future design. The operator of the centrifuge sets the percentage of the isotopes in the tails as well as that in the product. Given these percentages and the percentage in the feed one can use the Dirac value functions to evaluate the work needed to produce say a kilogram of uranium 235. The separative work unit is defined by the equation

$$SWU = WV(x_w) + PV(x_P) - FV(x_F).$$

(26)

Here the various $x_i$'s are the concentrations and the V's are the Dirac value functions. W,P and F are the quantities of waste, product and feed usually measured in kilograms. However what one does is to divide by P and write

$$\text{SWU/kilogram} = W/PV(x_w) + V(x_P) - F/PV(x_F).$$

(27)

In this process the quantity of uranium is preserved which means that these ratios are fixed by the concentrations. If you set the product to be say one kilogram you have

$$W/kg = (x_f - x_p)/(x_w - x_f)$$

(28)

$$F/kg = (x_w - x_p)/(x_w - x_f).$$

(29)

This means that the SWU-"Swoo"-per kilogram can be readily computed by using one of the many SWU calculators that you find on the web. Here are a few samples.[16]

|  | $x_f$ | $x_w$ | $x_p$ | SWU/kg |
|---|---|---|---|---|
| U-235 |  |  |  |  |
|  | 0.00711 | 0.0025 | 0.95 |  |
| 232.39 |  |  |  |  |
|  | 0.044 | 0.0025 | 0.95 |  |
| 72.46 |  |  |  |  |
|  | 0.199 | 0.0025 | 0.95 |  |
| 22.51 |  |  |  |  |

What strikes one is how rapidly the SWU fall off as the feed sample becomes more enriched. The first case is natural uranium. The second is reactor grade uranium and the third is the upper limit of what is called low enriched uranium. We can understand the trend if we imagine looking for needles in a haystack-the needles being U-235. The more highly enriched the feed the more "visible"are the needles and the easier our task is. To put these numbers in perspective, the Dirac limit for the kind of centrifuge that the Iranians have been employing is about five SWU per year although the actual SWU production might be at best half. The best modern centrifuges can produce over a hundred SWU per year. An implosion weapon needs about 20 kilograms. The implications for proliferation are clear. It is also instructive to say double the waste concentration to .005. The SWU requirement drops to 172.41 to produce highly enriched uranium from natural uranium. A homey illustration might help to illuminate the issues. Suppose we want to produce a certain amount of orange juice. If the price of oranges is not an issue we can leave more waste orange after each squeezing and use less energy per orange and use more oranges. When there is plenty of uranium hexafluoride available it might pay to increase the waste concentration. Again it was work of Dirac that led the way.

---

[16] These numbers are taken from a SWU calculator of R.L.Garwin. I thank him for making it available.

Coda:

Following his father's advice Dirac studied to be an engineer at Bristol University. He did well in his classes but badly when he went as an apprentice to the British Thompson Houston works in Rugby. After graduation as a consequence he could not find work as an engineer but was able to stay on at the university auditing mathematics classes until he was able to get a scholarship to Cambridge. One wonders what would have happened to twentieth century physics if Dirac had gotten a job as an engineer.

Appendix: This appendix presents a calculation of the quantity $\langle x_t | x_0 \rangle$ using path integrals in the simplest case possible where x propagates in time as a free particle. The calculation which I take from unpublished notes of M.Gell-Mann and M.L.Goldberger[17] already contains many features of the method. Doing realistic cases becomes very complicated.

---

[17] I thank Murph Goldberger for sharing these notes.

If the particle propagates as a free particle its Hamitonian is simply $p^2/2m$. But the action is written in terms of the coordinates so we solve the Heisenberg equation for $x_t$ to find

$$x_t = x_0 + p_0 t/m.$$

(1)

It will be useful to evaluate the commutator of $x_o$ and $x_t$. Using the previous equation

$$[x_0, x_t] = i\hbar t/m.$$

(2)

If we substitute the expression $p_0 = m/t(x_t - x_0)$ directly into the Hamiltonian we get terms involving both the product $x_0 x_t$ and $x_t x_0$ which are not the same. In the action we want to replace the operators by their eigen-values so we "well order" the Hamiltonian so that the $x_t$ terms are always to the left of the $x_o$ terms. Using the commutator the well-ordered Hamiltonian can be written as

$$H = m/2t^2 \{x_0^2 + x_t^2 - 2x_t x_0\} - i\hbar/2t.$$

(3)

Replacing these by their eigen-values in the equation for $S(x_t, x_o, t)$ we have the Hamilton-Jacobi equation

$$-\partial S/\partial t = (m/2t^2)(x_t - x_0)^2 - i\hbar/2t.$$

(4)

This equation has the solution

$$S(x', x) = (m/2t)(x'-x)^2 + i \ln((\alpha t)^{1/2}).$$

(5)

Here α is a constant that is determined by the condition that as t goes to zero $\exp(i/\hbar(S(x',x))) \to \delta(x'-x)$. There is no transition. We can now use the expression

$$\delta(x) = \lim_{\varepsilon \to 0} \{(1/\sqrt{2\pi i \varepsilon}) \exp(ix^2/2\varepsilon)\}$$

(6)

to evaluate α. This gives us

$$(x'|x) = \exp(im(x'-x)^2/2\hbar t)/(2\pi i\hbar t/m)^{1/2}.$$

(7)

This object is referred to as a "propagator" since it propagates the state forward in time.

$$|x'\rangle = \int dx \langle x'|x\rangle |x\rangle.$$

(8)